\documentstyle[12pt]{article}

\begin{document}

\setcounter{page}{1}

\pagestyle{plain} \vspace{1cm}
\begin{center}
\Large{\bf Cosmological braneworld solutions with bulk scalar field in DGP setup }\\
\small \vspace{1cm} {\bf Kourosh
Nozari\footnote{knozari@umz.ac.ir}}\quad and\quad {\bf M.
Khamesian\footnote{m.khamesian@umz.ac.ir}}\quad and
\quad{\bf N. Rashidi\footnote{n.rashidi@umz.ac.ir}}\\
\vspace{0.5cm} { Department of Physics, Faculty of Basic
Sciences,\\
University of Mazandaran,\\
P. O. Box 47416-95447, Babolsar, IRAN}

\end{center}

\vspace{1cm}
\begin{abstract}
We study cosmological dynamics of a canonical bulk scalar field in
the DGP setup within a superpotential approach. We show that the
normal branch of this DGP-inspired model realizes a late-time de
Sitter expansion on the brane. We extend this study to the case that
the bulk contains a phantom scalar field.  Our detailed study in the
supergravity-style analysis reveals some yet unexplored aspects of
cosmological dynamics of bulk scalar field in the normal DGP setup.
Some clarifying examples along with numerical analysis of the model
parameter space are presented in each case. \\
{\bf PACS}: 04.50.-h, \,98.80.Cq,\, 95.36.+x,\, 04.65.+e \\
{\bf Key Words}: Braneworld Cosmology, Induced Gravity, Bulk Scalar
Field, Supergravity and Superpotentials.
\end{abstract}
\newpage

\section{Introduction}
In the revolutionary braneworld viewpoint, our universe is a
$3$-brane embedded in an extra dimensional bulk. Standard matter and
all interactions are confined on the brane; only graviton and
possibly non-standard matter are free to probe the full bulk [1,2].
Based on the braneworld viewpoint, our universe may contain many
more dimensions than those we experience with our senses. The most
compelling reasons to believe in extra dimensions are that they
permit new connections between physical properties of the observed
universe and suggest the possibility for explaining some of its more
mysterious features. Extra dimensions can have novel implications
for the world we see, and they can explain phenomena that seem to be
mysterious when viewed from the perspective of a three-dimensional
observer. Even if one is doubtful about string theory due to, for
instance, its huge number of landscapes, recent researches have
provided perhaps the most compelling argument in the favor of extra
dimensions: a universe with extra dimensions might contain clues to
physics puzzles that have no convincing solutions without them. This
reason alone makes extra dimensional theories worthy of
investigation. In this streamline, the braneworld models that are
inspired by ideas from string theory provide a rich and interesting
phenomenology, where higher-dimensional gravity effects in the early
and late universe can be explored, and predictions can be made in
comparison with high-precision cosmological data.  Even for the
simplest models of RS and DGP, braneworld cosmology brings new
implications on the inflation and structure formation [1,3]. Also it
brings new ideas for dark energy and opens up exciting prospects for
subjecting M-theory ideas to the increasingly stringent tests
provided by high-precision astronomical observations [3]. At the
same time, braneworld models provide a rich playground for probing
the geometry and dynamics of the gravitational field and its
interaction with matter [3]. In these respects, the braneworld model
of Dvali, Gabadadze and Porrati (DGP) is a scenario that gravity is
altered at immense distances by the excruciatingly slow leakage of
gravity off our $3$-brane universe. In this braneworld scenario, the
bulk is considered as empty except for a cosmological constant and
the matter fields on the brane are considered as responsible for the
evolution on the brane [4,5]. The self-accelerating DGP branch
explains late-time speed-up by itself, without recourse to dark
energy or other mysterious components [5,6]. Even the normal DGP
branch has the potential to realize an effective phantom phase via
dynamical screening of the brane cosmological constant [7].

Here we are going to study cosmological dynamics in a DGP setup with
a bulk canonical/phantom scalar field. Many authors have studied the
cosmological consequences of a bulk scalar field (see for instance
[8,9,10]). One of the first motivations to introduce a bulk scalar
field was to stabilize the distance between the two branes in the
Randall-Sundrum two-brane model [11]. A second motivation for
studying scalar fields in the bulk is that such a setup could
provide some clue to solve the cosmological constant problem [12].
Models with inflation driven by bulk scalar field have been studied
and it is shown that inflation is possible without inflaton on the
brane [13]. Generally, the scalar field living in the bulk affects
the cosmological dynamics on the brane considerably. The evolution
of this field has some interesting cosmological implications; it can
give rise self-acceleration and phantom-like phase even in the
normal DGP branch of the model in some appropriate situations.
Solving the field equations for a braneworld cosmology with bulk
scalar field is not generally an easy task. Nevertheless, during the
past decade attempts have been performed to handle this problem. As
an attempt, one can express the five-dimensional Einstein equations
in terms of 4-dimensional tensors on the brane. Then, by using the
Darmois-Israel matching conditions, one can determine these tensors
[14]. However, this approach cannot determine all the tensors on the
brane. As has been pointed out in Ref. [15], one can proceed further
by making assumptions about the bulk solution. But, it is not
obvious that such assumptions are justified [9]. To determine which
solutions of the 4-dimensional Einstein equations are allowed, one
should solve the full 5-dimensional field equations. Full bulk
solutions for static braneworlds have been found in some special
cases (see [9] and references therein). If the scalar field
potential takes a supergravity-like form, the field equations can be
reduced to first order equations, which can be solved with relative
ease [16] (see also [9]). In this paper, we generalize the work of
Davis [9] to the DGP setup. We consider an extension of the DGP
scenario that the bulk is non-empty and contains a canonical or
phantom scalar field. Since the self-accelerating DGP branch has
ghost instabilities, we restrict our study to the normal DGP branch
of the model. The bulk equations of motion are derived and some
special classes of solutions are presented. We determine the
evolution of the brane when the potential of the scalar field takes
a supergravity-like form. Some clarifying examples along with
numerical analysis of the model parameter space are presented in
each case. The importance of this work lies in the fact that bulk
scalar field in DGP setup has not been studied in supergravity-style
analysis yet. Also our detailed study in this framework reveals some
yet unexplored aspects of cosmological dynamics of the bulk scalar
field in DGP setup.

\vspace{1cm}
\section{A canonical bulk scalar field in the DGP setup}
\subsection{The bulk field equations}

The five-dimensional action for a DGP-inspired braneworld model with
a bulk canonical scalar field can be written as follows
\begin{equation}
S=\int_{bulk}d^{5}x\sqrt{-g}\Big\{\frac{1}{2\kappa_{5}^{2}}\,
{^{(5)}}R-\frac{1}{2}(\nabla\phi)^{2}-V(\phi)\Big\}
-\int_{brane}d^{4}x\sqrt{-h}\Big(\frac{1}{2\kappa_{4}^{2}}R+\frac{1}{2\kappa_{5}^{2}}[K]+{\cal{L}}_{b}(\phi)\Big)\,,
\end{equation}
where $g_{AB}$ is the bulk metric and $h_{AB}$ is the induced metric
on the brane. They are related by $h_{AB}=g_{AB}-n_{A}n_{B}$, where
$n_{A}$ is the unit vector normal to the 3-brane and $A, B$ are the
five dimensional indices. The Gibbons-Hawking boundary term is
included via jump of the trace of the extrinsic curvature $[K]$\, in
the brane action.\, Also, $\kappa_{5}^2=\frac{8\pi}{M_{5}^3}$, where
$M_{5}$ is the fundamental five-dimensional Planck mass. The brane
Lagrangian ${\cal{L}}_{b}(\phi)$ includes all the Standard Model
fields which are confined to the brane, and depends on the bulk
scalar field. Varying the action with respect to the bulk scalar
field and also the bulk metric gives the bulk equations of motion
\begin{equation}
\nabla^{2}\phi=\frac{dV}{d\phi}+\frac{\sqrt{-h}}{\sqrt{-g}}\frac{d{\cal{L}}_{b}(\phi)}{d\phi}\delta(y)\,,
\end{equation}
\begin{equation}
G_{B}^{A}=\kappa_{5}^{2}\Bigg(\nabla^{A}\phi\nabla_{B}\phi-\delta_{B}^{A}\Big[\frac{1}{2}(\nabla\phi)^{2}
+V(\phi)\Big]\Bigg)+\delta(y)\kappa_{5}^{2}T_{B}^{A(brane)}\,,
\end{equation}
where
\begin{equation}
T^{(brane)}_{AB}=-2\frac{\delta {\cal{L}}_{brane}}{\delta
h^{AB}}+h_{AB}{\cal{L}}_{brane}.
\end{equation}
$T^{(brane)}_{AB}$ is the energy-momentum tensor localized on the
brane. $\delta(y)$ is the Dirac delta function with support on the
brane which we assume to be located at $y=0$ where $y$ is the
coordinate of the extra dimension. The action (1) implies the
following jump conditions
\begin{equation}
\Big[N^{A}\nabla_{A}\phi\Big]=\frac{\delta
{\cal{L}}_{b}(\phi)}{\delta \phi}\,,
\end{equation}
\begin{equation}
\Big[K_{AB}-Kh_{AB}\Big]=-\kappa_{4}^{2}T_{AB}^{(brane)}.
\end{equation}
The Gauss-Codacci equations relate projections of the bulk Einstein
tensor to the extrinsic curvature via
\begin{equation}
N^{A}G_{AB}h_{C}^{B}=D_{A}K_{C}^{A}-D_{C}K\,,
\end{equation}
where $D_{A}$ is the covariant derivative with respect to the bulk
metric. Combining this with (3) and substituting it into the jump
conditions (5) and (6) we find the effective energy-momentum
conservation equation on the brane
\begin{equation}
D_{A}T_{C}^{A(brane)}=-\frac{\delta
{\cal{L}}_{b}(\phi)}{\delta\phi}D_{C}\phi.
\end{equation}
To formulate cosmological dynamics on the brane, we assume the
following line element
\begin{equation}
ds^{2}=g_{AB}dx^{A}dx^{B}=-n^{2}(y,t)dt^{2}+a^{2}(y,t)\gamma_{ij}dx^{i}dx^{j}+b^{2}(y,t)dy^{2}
\end{equation}
where $\gamma_{ij}$ is a maximally symmetric 3-dimensional metric
defined as $\gamma_{ij}=\delta_{ij}+k\frac{x_{i}x_{j}}{1-kr^{2}}$
where $k=-1,0,+1$ parameterizes the spatial curvature and
$r^{2}=x_{i}x^{i}$.

Since we consider here homogeneous and isotropic geometries inside
the brane, $T_{B}^{A(brane)}$ can be expressed quite generally in
the following form
\begin{equation}
T_{B}^{A(brane)}=\frac{1}{b}diag(-\rho_{b}\,,\,p_{b}\,,\,p_{b}\,,\,p_{b}\,,\,0)\,.
\end{equation}
The extrinsic curvature tensor in the background metric (9) is given
by
\begin{equation}
K_{B}^{A}=diag\Big(\frac{n'}{nb}\,,\,\frac{a'}{ab}\delta_{j}^{i}\,,\,0\Big).
\end{equation}
So the jump conditions (5) and (6) are given as follows (we note
that in the forthcoming equations,
$\frac{\kappa_{5}^{2}}{2\kappa_{4}^{2}}\equiv r_{c}$ where $r_{c}$
is the DGP crossover scale)
\begin{equation}
\frac{[a']}{a_{0}b_{0}}=-\frac{\kappa_{5}^{2}}{3}\rho_{b}
+\frac{\kappa_{5}^{2}}{\kappa_{4}^{2}n_{0}^{2}}\Bigg\{\frac{\dot{a}_{0}^{2}}{a_{0}^{2}}+
k\frac{n_{0}^{2}}{a_{0}^{2}}\Bigg\}
\end{equation}
\begin{equation}
\frac{[n']}{n_{0}b_{0}}=\frac{\kappa_{5}^{2}}{3}(3p_{b}+2\rho_{b})+
\frac{\kappa_{5}^{2}}{\kappa_{4}^{2}n_{0}^{2}}\bigg\{-\frac{\dot{a}_{0}^{2}}{a_{0}^{2}}-
2\frac{\dot{a}_{0}\dot{n}_{0}}{a_{0}n_{0}}+\frac{2\ddot{a}_{0}}{a_{0}}-k\frac{n_{0}^{2}}{a_{0}^{2}}\Bigg\}
\end{equation}
\begin{equation}
\frac{[\phi']}{b_{0}}=\frac{\delta {\cal{L}}_{b}(\phi)}{\delta \phi}
\end{equation}
where a prime marks differentiation with respect to $y$ and a dot
denotes differentiation with respect to $t$. The subscript $0$ marks
quantities at $y=0$ (on the brane). Also $[A] = A(0^{+}) - A(0^{-})$
denotes the jump of the function $A$ across $y = 0$. Assuming
$Z_{2}$-symmetry about the brane for simplicity, the junction
conditions (12)-(14) can be used to compute $a'$, $n'$ and $\phi'$
on two sides of the brane. The energy-momentum conservation equation
(8) on the brane becomes
\begin{equation}
\dot{\rho}_{b}+3\frac{\dot{a}_{0}}{a_{0}}(\rho_{b}+p_{b})=\frac{\delta
{\cal{L}}_{b}(\phi)}{\delta \phi}\dot{\phi}_{0}.
\end{equation}
Because of the presence of the time-dependent bulk scalar field and
$\phi$-dependent couplings in the standard model lagrangian, the
right hand side of the above equation is non-zero and shows the
amount of energy non-conservation (due to bulk-brane energy-momentum
transfer) of the matter fields on the brane.

\subsection{DGP braneworld cosmology with a bulk scalar field}

We use the methods presented in Refs. [16,17] (see also [9]) to
obtain a special class of solutions for a DGP braneworld cosmology
with a bulk scalar field. In this respect, following [17], we
introduce the quantity $F$ as a function of\, $t$ and $y$ as follows
\begin{equation}
F(t,y)=-\Big(\frac{\dot{a}}{an}\Big)^{2}+\Big(\frac{a'}{ab}\Big)^{2}\,.
\end{equation}
So, the components of the Einstein tensor can be rewritten in the
following simple forms
\begin{equation}
G_{\,0}^{0}-\frac{\dot{a}}{a'}G_{\,5}^{0}=\frac{3}{2a^{3}a'}\partial_{y}(a^{4}F)-\frac{3k}{a^{2}},
\end{equation}
\begin{equation}
G_{\,5}^{5}-\frac{a'}{\dot{a}}G_{\,0}^{5}=\frac{3}{2a^{3}\dot{a}}\partial_{t}(a^{4}F)-\frac{3k}{a^{2}}.
\end{equation}\\

In the presence of the bulk scalar field, the left hand sides of
these two equations are not the same. But for special class of
solutions with $\phi=\phi(a)$, they are equivalent and in this case
$F=F(a)$. In this situation, both (17) and (18) then reduce to
\begin{equation}
\kappa_{5}^{2}V(\phi)+\frac{\kappa_{5}^{2}}{2}
F\Big(a\frac{d\phi}{da}\Big)^{2}+\Bigg\{\frac{3\kappa_{5}^{2}}{\kappa_{4}^{2}}\bigg(\frac{\dot{a}}{a}\bigg)^{2}
+\frac{3\kappa_{5}^{2}}{\kappa_{4}^{2}}\bigg(\frac{k}{a^{2}}\bigg)+\kappa_{5}^{2}\,\rho_{b}
\Bigg\}\delta(y)+6F +\frac{3}{2}a\frac{dF}{da}-\frac{3k}{a^{2}}=0.
\end{equation}
We choose a Gaussian normal coordinate system so that
$b^{2}(y,t)=1$. Also we assume that $t$ as a proper cosmological
time on the brane has scaled so that $n_{0}=1$. By adopting a
$Z_{2}$ symmetry across the brane, equations (13) and (16) yield the
following generalization of the Friedmann equation for cosmological
dynamics on the DGP brane
\begin{equation}
\Big(\frac{\dot{a}_{0}}{a_{0}}\Big)^{2}=\frac{1}{3}\kappa_{4}^{2}\rho_{b}+
\frac{2\kappa_{4}^{4}}{\kappa_{5}^{4}}-\frac{k}{a^{2}}-
\frac{2\kappa_{4}^{2}}{\kappa_{5}^{2}}\sqrt{\frac{\kappa_{4}^{4}}
{\kappa_{5}^{4}}+\frac{1}{3}\kappa_{4}^{2}\rho_{b}-
\frac{k}{a^{2}}+F_{0}}.
\end{equation}
We note that here we consider only the normal, ghost-free branch of
the solutions. Since we consider $\phi=\phi(a)$, the field equation
(2) reduces to
\begin{equation}
\nabla^{2}\phi=a^{2}F\Big(\frac{d^{2}\phi}{da^{2}}+\frac{1}{a}\frac{d\phi}{da}\Big)
+\Big[(G^{0}_{\,0}+G_{\,5}^{5})\frac{a}{3}+\frac{2k}{a}\Big]\frac{d\phi}{da}=
\frac{d V}{d\phi}+\delta(y)\frac{\sqrt{-h}}{\sqrt{-g}}\frac{\delta
{\cal{L}}_{b}(\phi)}{\delta\phi}.
\end{equation}
Substituting equations $G_{AB}=\kappa^2 T_{AB}$ and (19) into (21),
we find
$$F\Big(a\frac{d}{da}\Big)^{2}\phi+\Bigg\{\frac{2a}{3}\bigg[\frac{\kappa_{5}^{2}}{2}
F\Big(a\frac{d\phi}{da}\Big)^{2}+\Big\{\frac{3\kappa_{5}^{2}}{\kappa_{4}^{2}}\Big(\frac{\dot{a}}{a}\Big)^{2}
+\frac{3\kappa_{5}^{2}}{\kappa_{4}^{2}}\Big(\frac{k}{a^{2}}\Big)
+\kappa_{5}^{2}\,\rho_{b} \Big\}\delta(y)+6F
+\frac{3}{2}a\frac{dF}{da}\bigg]-\frac{k}{a}\Bigg\}\frac{d\phi}{da}$$
\begin{equation}
-\frac{d V}{d\phi}-\delta(y)\frac{\sqrt{-h}}{\sqrt{-g}}\frac{\delta
{\cal{L}}_{b}(\phi)}{\delta\phi}=0.
\end{equation}
Thus the original partial differential field equations have been
reduced to an ordinary differential equation.

\subsection{Supergravity-style solutions}

In order to generate some solutions of the field equations, we
introduce a special supergravity-style potential, $V(\phi)$, as
follows [9]
\begin{equation}
V(\phi)=\frac{1}{8}\Big(\frac{dW}{d\phi}\Big)^{2}-\frac{\kappa_{5}^{2}}{6}W^{2}.
\end{equation}
Assuming $k=0$, the field equations (19) and (22) are satisfied if
\begin{equation}
F=\frac{\kappa_{5}^{4}}{36}W^{2},
\end{equation}
\begin{equation}
a\frac{d\phi}{da}=-\frac{3}{\kappa_{5}^{2}W}\frac{dW}{d\phi}.
\end{equation}
Here, $W$ is referred to as a superpotential, but we note that
supergravity is not required for solutions (23)-(25) to satisfy the
equations (19) and (22). Now we can rewrite the Friedmann equation
(20) in terms of $W$ as follows
\begin{equation}
\frac{\dot{a}^{2}_{0}}{a_{0}^{2}}=\frac{1}{3}\kappa_{4}^{2}\rho_{b}+\frac{2\kappa_{4}^{4}}{\kappa_{5}^{4}}
-\frac{2\kappa_{4}^{2}}{\kappa_{5}^{2}}\sqrt{\frac{\kappa_{4}^{4}}{\kappa_{5}^{4}}
+\frac{1}{3}\kappa_{4}^{2}\rho_{b}+\frac{\kappa_{5}^{4}}{36}W_{0}^{2}}.
\end{equation}

From equations (25) and (26) we find the the time variation of the
scalar field on the brane as
\begin{equation}
\dot{\phi}^{2}=\Bigg[\frac{1}{3}\kappa_{4}^{2}\rho_{b}+\frac{2\kappa_{4}^{4}}{\kappa_{5}^{4}}-
\frac{2\kappa_{4}^{2}}{\kappa_{5}^{2}}\sqrt{\frac{\kappa_{4}^{4}}{\kappa_{5}^{4}}+\frac{1}{3}\kappa_{4}^{2}\rho_{b}
+\frac{\kappa_{5}^{4}}{36}W_{0}^{2}}\Bigg]\Bigg(\frac{9}{\kappa_{5}^{4}W^{2}_{0}}\Bigg)
\Bigg(\frac{dW}{d\phi}\Bigg)_{0}^{2}.
\end{equation}
The jump conditions (12) and (14) are related via equation (25).
Consistency of jump conditions for a ${Z}_{2}$-symmetric DGP brane
is guaranteed if
\begin{equation} \frac{\delta
{\cal{L}}_{b}(\phi)}{\delta\phi}=\Bigg(-\frac{\kappa_{5}^{2}}{6}\rho_{b}
+\frac{\kappa_{5}^{2}}{2\kappa_{4}^{2}}\frac{\dot{a}^{2}_{0}}{a_{0}^{2}}\Bigg)
\Bigg(-\frac{6}{\kappa_{5}^{2}W}\frac{dW}{d\phi}\Bigg)_{0}.
\end{equation}\\
So, the energy on the DGP brane is not conserved in this setup.
Combining equations (15) and (28), one finds the energy conservation
equation in terms of $W$ as follows

\begin{equation}
\dot{\rho}_{b}+3\frac{\dot{a}_{0}}{a_{0}}\big(\rho_{b}+p_{b}\big)=\Bigg[\rho_{b}-\frac{3}{\kappa_{4}^{2}}
\frac{\dot{a}^{2}_{0}}{a_{0}^{2}}\Bigg]\frac{\dot{W}_{0}}{W_{0}}\,
\end{equation}
We note that this equation in the case of RSII braneworld scenario
has a simpler structure (see for instance, equation (4.7) of Ref.
[9]). In our case, due to the presence of two terms on the right
hand side of equation (29), there will be new possibilities with
different cosmological implications. To discuss the status of the
conservation equation in different cases, we define the parameter
$X$ so that $X\equiv \frac{3}{\kappa_{4}^{2}}
\frac{\dot{a}^{2}_{0}}{a_{0}^{2}}$ for simplicity.\\

There are three possibilities as follows:\\

\textbf{\textit{A: $\rho_{b}>X$}}\\

For negative values of $\frac{\dot{W}_{0}}{W_{0}}$, energy will leak
off the brane. For positive values of this quantity, energy will
flow from the bulk into the brane.\\

\textbf{\textit{B: $\rho_{b}<X$}}\\

In this case for positive values of $\frac{\dot{W}_{0}}{W_{0}}$,
energy will leak off the brane and for negative values, energy will
transfer from the bulk into the brane.\\

\textbf{\textit{C: $\rho_{b}=X$}}\\

In this case the right hand side of equation (29) vanishes and the
sign of $\frac{\dot{W}_{0}}{W_{0}}$ is not important, thus the
energy on the brane is conserved.\\

We note that for all mentioned cases, if
$\frac{\dot{W}_{0}}{W_{0}}=0$, we have conservation of energy on the
brane. Now we consider the simplest generalization of the brane
energy density as
\begin{equation}
\rho_{b}=W_{0}\rho
\end{equation}
where $\rho$ is proportional to the energy density of the ordinary
matter on the brane. This generalization has its origin in the fact
that matter Lagrangian on the brane, that is
${\cal{L}}_{b}(\phi)$,\, depends on the bulk scalar field, $\phi$.
The effective Friedmann equation (26) then becomes
\begin{equation}
\frac{\dot{a}^{2}_{0}}{a_{0}^{2}}=\frac{1}{3}\kappa_{4}^{2}W_{0}\,\rho+\frac{2\kappa_{4}^{4}}{\kappa_{5}^{4}}
-\frac{2\kappa_{4}^{2}}{\kappa_{5}^{2}}\sqrt{\frac{\kappa_{4}^{4}}{\kappa_{5}^{4}}
+\frac{1}{3}\kappa_{4}^{2}W_{0}\,\rho+\frac{\kappa_{5}^{4}}{36}W_{0}^{2}}.
\end{equation}
This equation implies that if $W_{0}$ tends to $\infty$ at late
time, the cosmological evolution on the brane is not generally
compatible with observations. On the other hand, for $W_{0}=0$,\,
the result is not compatible with observation too. Only for a
constant $W_{0}$ we find a viable DGP-like cosmology in this simple
generalization. In this case one obtains a self-accelerating
solution which explains the late-time cosmic speed up. We note that
a phantom-like prescription can be realized in this case in the same
way as has been shown in Ref. [18] a canonical scalar field on the
brane.

In the next section, we will consider some specific examples of
superpotential and we will discuss their cosmological implications.
Specifically, we will consider the evolution of $W$ in order to
study the status of the conservation equation.

\subsection{Some specific examples}

We consider the following exponential form of the superpotential [9]
\begin{equation}
W=c\Bigg[\frac{e^{-{\alpha}_{1}{\phi}}}{{\alpha}_{1}}+{s}\frac{e^{{\alpha}_{2}{\phi}}}{{\alpha}_{2}}\Bigg] ,
\end{equation}
where $s=\pm1$ and $\alpha_1\geq\left|\alpha_2\right|$. For
$\alpha_2=0$, we use the following form of the superpotential
\begin{equation}
W=c\Bigg[\frac{e^{-{\alpha}_{1}{\phi}}}{{\alpha}_{1}}+\frac{s}{\kappa_{5}}\Bigg]
.
\end{equation}
The corresponding potentials obtained from (23) are
\begin{equation}
V=\frac{c^{2}}{8}\Bigg[\Bigg(1-\frac{4{\kappa_{5}}^{2}}{3{\alpha}_{1}^2}\Bigg)e^{-2{\alpha}_{1}{\phi}}+
\Bigg(1-\frac{4{\kappa_{5}}^{2}}{3{\alpha}_{2}^{2}}\Bigg)e^{2{\alpha}_{2}{\phi}}-2{s}
\Bigg(1+\frac{4{\kappa_{5}}^{2}}{3{\alpha}_1{\alpha}_{2}}\Bigg)e^{({\alpha}_{2}-{\alpha}_{1}){\phi}}\Bigg]
.
\end{equation}
and (for ${\alpha}_{2}=0$)
\begin{equation}
V=\frac{c^{2}}{8}\Bigg[\Bigg(1-\frac{4{\kappa_{5}}^{2}}{3{\alpha}_{1}^2}\Bigg)e^{-2{\alpha}_{1}{\phi}}-2{s}
\frac{4{\kappa_{5}}}{3{\alpha}_{1}}e^{-{\alpha}_{1}{\phi}}-\frac{4}{3}\Bigg]
.
\end{equation}
For $V$ bounded from below, only some values of the parameters are
allowed [9]. For $W$ as given by (32), one can solve equation (25)
to find
\begin{equation}
\ln(\frac{a}{a_{*}})=-\frac{{\kappa_{5}}^2}{3{\alpha}_{1}{\alpha}_{2}}
\ln\left|e^{{\alpha}_{1}{\phi}}-se^{-{\alpha}_{2}{\phi}}\right| ,
\end{equation}
where $a_*$ is an arbitrary constant. For ${\alpha}_{2}=0$, equation
(25) is solved by
\begin{equation}
\ln{\big(\frac{a}{{a}_*}\big)}=\frac{\kappa_{5}}{3{\alpha}_{1}}\big({\kappa_{5}}{\phi}+{s}e^{{\alpha}_{1}{\phi}}\big).
\end{equation}
We emphasize here that although the results of this section
(equations (36) and (37)) seems to be formally the same as the
results obtained in [9] for Randall-Sundrumm II (RSII) braneworld,
but one should remember that our Friedmann equation (26) and hence
the Hubble parameter and scale factor differ from corresponding
quantities in Ref. [9]. With these new quantities, equations (36)
and (37) differs essentially from corresponding equations (5.5) and
(5.6) of Ref. [9] for RSII case.\\

Depending on the choice of $s$ and the sign of ${\alpha}_{2}$, there
will be a variety of cosmological evolution on the brane with
several interesting implications. To illustrate further, in which
follows, we will separate each of these subcases for some values of
$\alpha_{1}$ and $\alpha_{2}$. Then we will consider the evolution
of scalar field versus the scale factor in normal DGP branch of the
scenario.\\

\textbf{2.4.1\quad ${\alpha}_{2}>0$, ${s}=+1$}\\

In this case, as $a$ goes from $0$ to ${\infty}$, scalar field rolls
down from either $+{\infty}$ or $-{\infty}$ to $0$. In figure 1, we
show the evolution of the scalar field versus the scale factor
explicitly. We fix $\alpha_{1}$ in a constant value and choose three
different values for $\alpha_{2}$. As we see, when the value of
$\alpha_{2}$ grows, the slope of the curve increases. This increase
means that for larger $\alpha_{2}$, the scalar field varies faster.
Also, for negative $\phi$, variation of $\alpha_{2}$ has no
significant effect on the slope of the curves since all curves
coincide in this case. Now we look at the status of the continuity
equation in this case. \\

\begin{figure}[htp]
\begin{center}\includegraphics{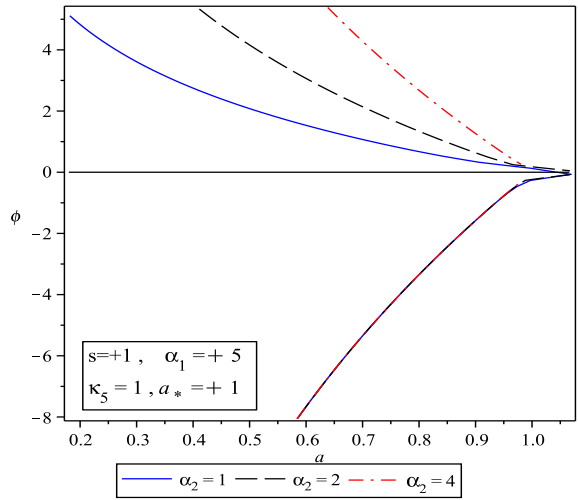} \vspace{6.5cm}
\end{center}
\caption{\small {Evolution of the scalar field with respect to the
scale factor for the case ${\alpha}_{2}>0$, ${s}=+1$. As the scale
factor evolves from $0$ to ${\infty}$, scalar field rolls down from
either $+{\infty}$ or $-{\infty}$ to $0$.}}
\end{figure}

$\clubsuit$ If we consider the case that $\phi$ starts from
$-\infty$, then as $\phi$ goes to $0$ (and $a$ increases),
superpotential evolves from infinity to a constant value with
$\dot{W}=0$ (see figure 5). During this evolution,
$\frac{\dot{W}}{W}<0$, and therefore from (29) we will have the
following conditions:\\

\textit{\textbf{A}:}  If $\rho_{b}>X$, the right hand side of
equation (29) becomes negative and this indicates that energy leaks
off the brane as scale factor increases and the universe expands.
This situation continues until $\phi$ and $\dot{W}$ tend to zero.
Then, there will be no energy leakage off the brane.\\

\textit{\textbf{B}:}  If $\rho_{b}<X$, the right hand side of
equation (29) becomes positive and the energy is sucked onto the
brane as universe expands. This situation continues until constant
value of $W$ ($\dot{W}=0$) in $\phi=0$. Then, there will be no
energy suction onto the brane.\\

$\clubsuit$ On the other hand, if we consider the case that $\phi$
starts from $+\infty$ and tends to zero as $a$ goes to infinity, the
superpotential evolves from $+\infty$ to a constant value where
$\dot{W}= 0$. During this stage, $\frac{\dot{W}}{W}>0$. In analogy
to the previous case, there are two possibilities:\\

\textit{\textbf{A}:} If $\rho_{b}<X$, the right hand side of
equation (29) becomes negative. This means that as the universe
expands, energy leaks off the brane until $\dot{W}=0$.\\

\textit{\textbf{B}:} If $\rho_{b}>X$, the right hand side of the
equation of continuity becomes positive and the energy sucks onto
the brane as scale factor increases. As soon as $\dot{W}$ vanishes,
the energy suction stops. \\

\textbf{2.4.2\quad ${\alpha}_{2}>0$, ${s}=-1$}\\

As can be seen from figure 2, for this choice of parameters,
${\phi}$ evolves from $+{\infty}$ to $-{\infty}$ or vice versa.
During this evolution, the scale factor varies from zero to a
maximum value and then reduces again to zero. For a constant value
of $\alpha_{1}$, increasing the value of $\alpha_{2}$ leads to
increasing in the slope of the curves when $\phi$ is positive. For
negative $\phi$, corresponding changes are not significant.
Regarding to the continuity equation we have the following
possibilities: \\

\begin{figure}[htp]
\begin{center}\includegraphics{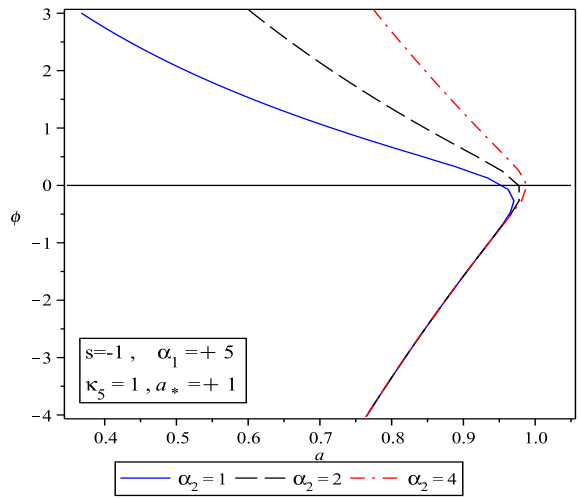} \vspace{6.5cm}
\end{center}
\caption{\small { Evolution of the scalar field with respect to the
scale factor for the case ${\alpha}_{2}>0$, ${s}=-1$. ${\phi}$ can
evolve from $+{\infty}$ to $-{\infty}$ or vice versa. During this
evolution, the scale factor changes from zero to a maximum value and
then reduces again to zero. }}
\end{figure}

$\clubsuit$ Equation (32) shows that at first, when $\phi$ is
$+\infty$ or $-\infty$, $W$ is $-\infty$ or $+\infty$ respectively
(see also figure 5). When $\phi$ starts to decrease from infinity,
we have $\frac{\dot{W}}{W}>0$. This situation continues until $\phi$
tends to a special value at which scale factor reaches its maximum
value and $\dot{a}=0$. This implies that:\\

\textit{\textbf{A}:}  For $\rho_{b}<X$, energy leaks off the brane
as scale factor increases and the universe expands. This leakage of
energy continues until the scale factor reaches its maximum value
and $\dot{a}=0$ (see figure 2). At this point, $W$ vanishes and
there is nothing left on the brane.\\

\textit{\textbf{B}:}  If $\rho_{b}>X$, the situation is very
different. When $\phi$ is $+\infty$ and $\frac{\dot{W}}{W}>0$, the
right hand side of equation (29) is positive. This means that energy
is sucked onto the brane whereas the universe expands. This suction
persists until scale factor reaches its maximum value.\\

$\clubsuit$ After scale factor reaches its maximum value, the
universe begins to re-collapse and scale factor decreases then. In
this stage as $\phi$ and $W$ goes to $-\infty$ and $+\infty$
respectively, $\frac{\dot{W}}{W}<0$. This continues until the scale
factor returns to zero where the superpotential tends to infinity.\\

\textit{\textbf{A}:} If $\rho_{b}<X$, the right hand side of
equation (29) becomes positive. This indicates that as the scale
factor decreases and the universe re-collapses, energy is sucked
onto the brane. This suction continues until the scale factor tends
to zero where $W$ becomes infinity.\\

\textit{\textbf{B}:} If $\rho_{b}>X$, the energy leaks off the brane
until the scale factor returns to zero where the
superpotential tends to infinity.\\

$\clubsuit$ When $\phi$ starts to increase from $-\infty$, we have
$\frac{\dot{W}}{W}<0$. This situation continues until $\phi$ tends
to a special value at which scale factor reaches its maximum value
and $\dot{a}=0$. This implies that:\\

\textit{\textbf{A}:}  For $\rho_{b}>X$, energy leaks off the brane
as scale factor increases and the universe expands. This leakage of
energy continues until the scale factor reaches its maximum value
and $\dot{a}=0$. At this point, $W$ vanishes.\\

\textit{\textbf{B}:}  If $\rho_{b}<X$, the right hand side of
equation (29) is positive. This means that energy is sucked onto the
brane whereas the universe expands. This suction persists until
scale factor reaches its maximum value.\\

$\clubsuit$ After scale factor reaches its maximum value, the
universe begins to re-collapse and scale factor decreases then. In
this stage as $\phi$ goes to $+\infty$, $\frac{\dot{W}}{W}>0$. This
continues until the scale factor returns to zero where the scalar
field tends to infinity. So:\\

\textit{\textbf{A}:} If $\rho_{b}>X$, the right hand side of
equation (29) becomes positive. This implies that as the scale
factor decreases and the universe re-collapses, energy is sucked
onto the brane. This suction continues until scale factor tends to
zero where $\phi$ becomes infinity.\\

\textit{\textbf{B}:} If $\rho_{b}<X$, the energy leaks off the brane
until the scale factor returns to zero where the scalar field tends to infinity.\\

\textbf{2.4.3\quad ${\alpha}_{2}<0$, ${s}=-1$}\\

In this case, the scalar field evolves from $-\infty$ to $+\infty$,
as the scale factor goes from zero to $\infty$. Figure 3 shows this
behavior. One can see from this figure that by decreasing the value
of $\alpha_{2}$, the slope of curves increases for positive scalar
field. However, for negative scalar field this change is not
significant and there is no considerable shift in curves. Regarding
to the continuity equation, the following issues are in order: \\

\begin{figure}[htp]
\begin{center}\includegraphics{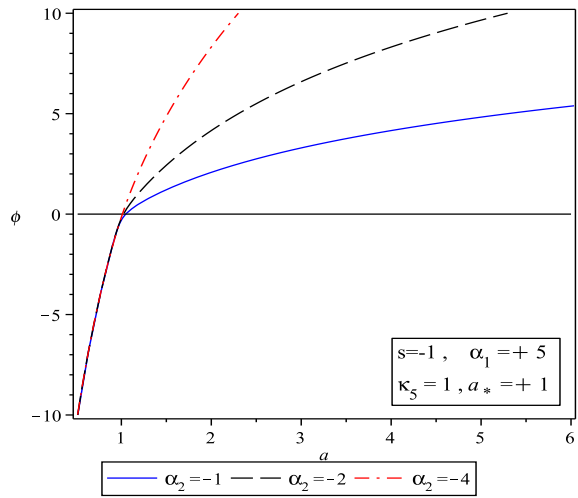} \vspace{6.5cm}
\end{center}
\caption{\small { Evolution of the scalar field with respect to the
scale factor for the case  with ${\alpha}_{2}<0$ and ${s}=-1$. As
scale factor goes from zero to $\infty$, the scalar field evolves
from $-\infty$ to $+\infty$.}}
\end{figure}

$\clubsuit$  At first, when the scalar field is $-\infty$, the
superpotential is $+\infty$. Then, by increasing the values of scale
factor and scalar field, the superpotential decreases to zero as
$\phi\rightarrow\infty$. During this evolution, we have
$\frac{\dot{W}}{W}<0$.\\

\textit{\textbf{A}:}  If $\rho_{b}>X$, as scale factor increases and
the univers expands, energy leaks off the brane and this leakage
continues until $W$ tends to zero at late time.\\

\textit{\textbf{B}:} If $\rho_{b}<X$, from equation (29) we find
that as the universe expands, the energy is sucked onto the brane.

\vspace{2cm}

\textbf{2.4.4\quad ${\alpha}_{2}<0$, ${s}=+1$}\\

For this choice of parameters, the solution starts from
$\phi=-\infty$ and $a=0$. The scalar field increases with scale
factor until some value of $a$ where $\dot{a}=0$. At this point $a$
reaches a relative maximum. After that, the scale factor begins to
decrease and scalar field continues its growing until it tends to
zero. At this point scale factor reaches its relative minimum, where
$\dot{a}$ is zero. Then, $\phi$ grows again with scale factor
towards infinity. This behavior can be seen in figure 4. Further,
this figure shows also that decreasing the value of $\alpha_{2}$
leads to increasing in the slope of curves for positive $\phi$. For
negative $\phi$, change in the values that $\alpha_{2}$ attains has
no significant effect on the evolution of $\phi$. Once again,
regarding to energy conservation in this case we arrive at the
following points:\\

\begin{figure}[htp]
\begin{center}\includegraphics{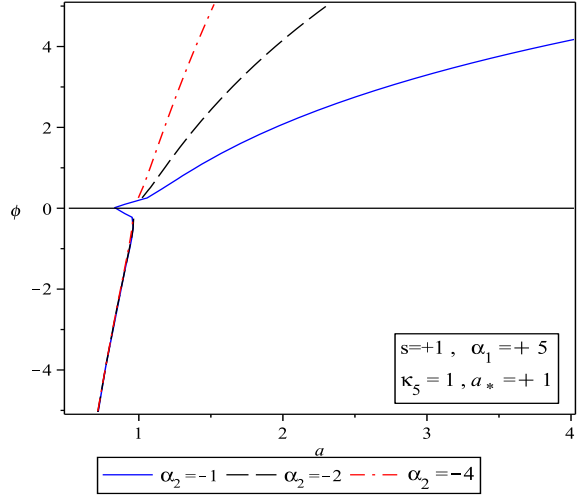} \vspace{6.5cm}
\end{center}
\caption{\small {Evolution of the scalar field with respect to the
scale factor for the case with ${\alpha}_{2}>0$ and ${s}=+1$. The
scalar field increases with scale factor until $a$ reaches a
relative maximum. After that, the scale factor begins to decrease
and the scalar field continues its growing until it tends to zero.
Then, $\phi$ grows to infinity when $a$ tends to infinity.}}
\end{figure}

$\clubsuit$  Firstly, when $\phi$ is $-\infty$, the superpotential
is $+\infty$. As $\phi$ increases and $a$ goes to its relative
maximum, $W$ tends to zero. During this stage,
$\frac{\dot{W}}{W}<0$. So we have:\\

\textit{\textbf{A}:} If $\rho_{b}>X$, the right hand side of
conservation equation (29) becomes negative and energy leaks off the
brane with expansion of the universe. This feature continues until
$W$ vanishes at the relative maximum of the scale factor.\\

\textit{\textbf{B}:} If $\rho_{b}<X$, the right hand side of
equation (29) becomes positive and energy is sucked onto the brane
as scale factor increases. This situation continues until the scale
factor reaches its relative maximum.\\

$\clubsuit$ Secondly, when $a$ decreases to its relative minimum and
$\phi$ tends to zero, the superpotential reaches its minimum (here
it is a negative value) and $\dot{W}$ vanishes. During this
stage, $\frac{\dot{W}}{W}>0$. We find that:\\

\textit{\textbf{A}:} If $\rho_{b}>X$, the right hand side of
equation (29) becomes positive. As the scale factor decreases and
the universe re-collapses, energy is sucked onto the brane until
$\phi=0$.\\

\textit{\textbf{B}:} If $\rho_{b}<X$, the right hand side of
conservation equation becomes negative. So, the energy leaks off the
brane and this leakage persists until the minimum value of
superpotential is achieved.\\

$\clubsuit$ Thirdly, when $a$ starts to increase from its relative
minimum towards infinity, the superpotential changes from its
minimum to zero when $\phi$ goes to $+\infty$ . During this
evolution, we have $\frac{\dot{W}}{W}<0$. Here also we have:\\

\textit{\textbf{A}:} If $\rho_{b}>X$, the right hand side of
equation (29) becomes negative and as $a$ and $\phi$ tend to
infinity, the energy leaks off the brane.\\

\textit{\textbf{B}:} If $\rho_{b}<X$, the right hand side of
equation (29) becomes positive. Here, as universe expands, the
energy is sucked onto the brane.\\

\begin{figure}[htp]
\begin{center}\includegraphics{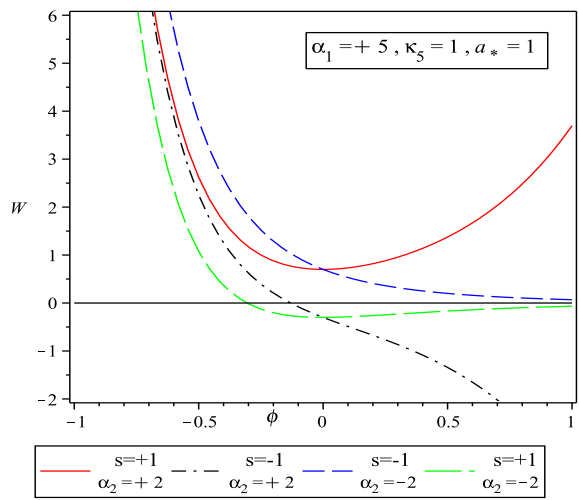} \vspace{6.5cm}
\end{center}
\caption{\small {Evolution of the superpotential versus the scalar
field. This evolution plays important role in the status of the
energy conservation.}}
\end{figure}
\vspace{2cm}
\textbf{2.4.5\quad ${\alpha}_{2}=0$, ${s}=\pm1$}\\

In these cases, for $s=+1$, scalar field evolves from $-\infty$ to
$+\infty$,\, while for $s=-1$ it evolves from $-\infty$ to $+\infty$
or vice versa. The value of $s$ determines status of this
evolution.\\

\begin{figure}[htp]
\begin{center}\includegraphics{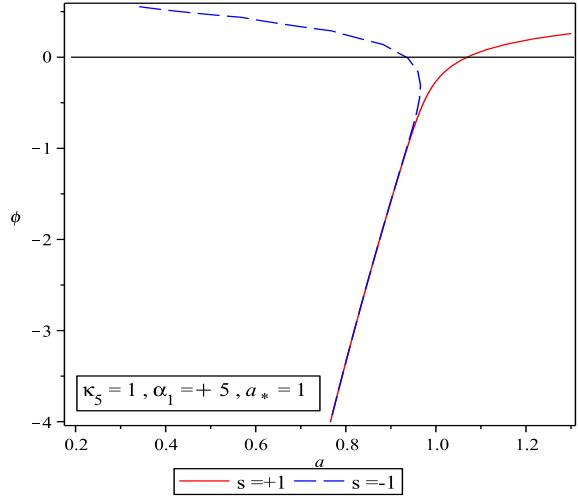} \vspace{6.5cm}
\end{center}
\caption{\small {Evolution of scalar field respect to scale factor.
If $s=-1$, scale factor has a maximum. But for $s=+1$ it continues
to infinity.}}
\end{figure}

$\clubsuit$ If $s=+1$, $\phi$ starts from $-\infty$ at $a=0$ and
continues to $+\infty$ as $a$ goes to infinity. When $\phi$ is
$-\infty$, the superpotential is $+\infty$. As scalar field
increases, the superpotential decreases to a constant value when
$\phi$ tends to infinity. During this stage $\frac{\dot{W}}{W}<0$.
So we have:\\

\textit{\textbf{A}:} If $\rho_{b}>X$, as $a$ and $\phi$ go to
infinity, the energy leaks off the brane. This leakage stops when
$W$ reaches a constant value.\\

\textit{\textbf{B}:} If $\rho_{b}<X$, the right hand side of
equation (29) becomes positive. Here, as universe expands, the
energy is sucked onto the brane.\\

$\clubsuit$ If $s=-1$, $\phi$ starts from $-\infty$ at $a=0$ and
increases until $a$ reaches its maximum. At this point $\dot{a}$ and
$W$ tend to zero. In this situation $\frac{\dot{W}}{W}<0$ and we
have:\\

\textit{\textbf{A}:} If $\rho_{b}>X$, as $a$ and $\phi$ go to
infinity the energy leaks off the brane. This leakage continues
until $W$ tends to zero.\\

\textit{\textbf{B}:} If $\rho_{b}<X$, as universe expands, the
energy is sucked onto the brane.\\

Just after $\dot{a}$ becomes zero, the scale factor starts to
decrease and the universe re-collapses. When $a$ goes from maximum
to zero, the scalar field continues its growth and tends to
$+\infty$. During this stage, $W$ decreases and we have
$\frac{\dot{W}}{W}>0$. So the following subcases are in order:\\

\textit{\textbf{A}:} If $\rho_{b}<X$, as the universe re-collapses,
the energy is leaks off the brane until $W$ tends to a constant
value.\\

\textit{\textbf{B}:} If $\rho_{b}>X$ as $a$ goes to zero, the energy
is sucked onto the brane. This suction persists until the scalar
field reaches $+\infty$.\\

$\clubsuit$ For $s=-1$, there is another situation that $\phi$
starts from $+\infty$ at $a=0$ and decrease until $a$ reaches its
maximum. At this point $\dot{a}$ and $W$ tend to zero. In this
situation $\frac{\dot{W}}{W}>0$ and we have:\\

\textit{\textbf{A}:} If $\rho_{b}<X$, as $a$ increase and $\phi$
decrease, the energy leaks off the brane. This leakage continues
until $a$ reaches its maximum value.\\

\textit{\textbf{B}:} If $\rho_{b}>X$, as universe expands, the
energy is sucked onto the brane.\\

Just after $\dot{a}$ becomes zero, the scale factor starts to
decrease and the universe re-collapses. When $a$ goes from maximum
to zero, the scalar field continues its decrease and tends to
$-\infty$. During this stage, $W$ increases and we have
$\frac{\dot{W}}{W}<0$. So the following subcases are in order:\\

\textit{\textbf{A}:} If $\rho_{b}>X$, as the universe re-collapses,
the energy is leaks off the brane until $W$ tends to a constant
value.\\

\textit{\textbf{B}:} If $\rho_{b}<X$ as $a$ goes to zero, the energy
is sucked onto the brane. This suction persists until the scalar
field reaches $+\infty$.\\

\begin{figure}[htp]
\begin{center}\includegraphics{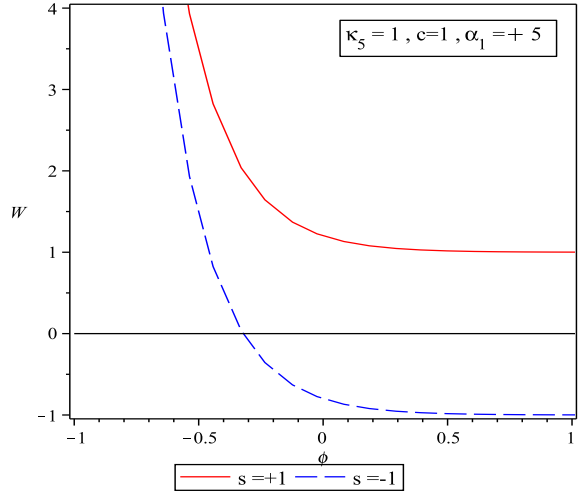} \vspace{6.5cm}
\end{center}
\caption{\small {Evolution of the superpotential versus the scalar
field for $\alpha_{2}=0$. If $s=-1$, the superpotential will pass
through a zero and then tends to a constant . If $s=+1$, the
superpotential has no root.}}
\end{figure}

In summary, assuming a simplest generalization of the brane energy
density $(\rho_{b}=W_{0}\rho)$,\, implies that if $W_{0}$ tends to
$+\infty$ at late time, the cosmological evolution on the brane is
not generally compatible with observations. On the other hand, for
$W_{0}=0$, the result is not compatible with observation too. Only
for a constant $W_{0}$ we find a viable DGP-like cosmology in this
simple generalization. So, we found in this analysis that the case
with $\alpha_{2}>0$ and $s=+1$ gives a viable DGP-like cosmology. We
should note also that in this case, the situation that the energy
leaks of the brane during universe expansion is more reliable.

After a detailed discussion on the energy conservation and possible
cosmological dynamics on the brane, in the next subsection we
determine cosmological dynamics in the bulk.

\subsection{Bulk solutions}

By solving the non-zero off-diagonal components of the Einstein
field equations, we are able to determine the $y$-dependence of the
metric or the scalar field ${\phi}$ in the bulk. To do this end, we
assume the fifth dimension is static. In other words, $\dot{b}=0$
and so we can adopt the gauge $b=1$. When ${\phi}={\phi}(a)$, we
have
\begin{equation}
{G}_{05}=3\Bigg(\frac{n'\dot{a}}{na}+\frac{\dot{b}{a'}}{ba}-\frac{\dot{a}'}{a}\Bigg)={\kappa}_{5}^{2}{T}_{05}=
{\kappa}_{5}^{2}{a'}\dot{a}\Bigg(\frac{d{\phi}}{da}\Bigg)^{2} .
\end{equation}
Then $n$ can be expressed in terms of $a$ according to the following
relation
\begin{equation}
\frac{\dot{a}}{n}=\beta(t)\frac{{\kappa}_{5}^{2}}{6}
\exp{\Bigg[-\frac{{\kappa}_{5}{^2}}{3}\int{a\Bigg(\frac{d{\phi}}{da}\Bigg)^2{da}}\Bigg]}
\end{equation}
where $\beta$ is a function of time alone and has no dependence on
$y$. If $V$ has a supergravity-like form as $(23)$, we can rewrite
the above expression as
\begin{equation}
\frac{\dot{a}}{n}=\beta(t)\frac{{\kappa}_{5}{^2}}{6}W({\phi}) .
\end{equation}
We set ${n}_{0}=1$, then  $(26)$ implies that
\begin{equation}
\beta=\frac{a_{0}}{W_{0}}\Bigg[\frac{72\kappa_{4}^{4}}{\kappa_{5}^{8}}+\frac{12\rho_{b}\kappa_{4}^{2}}{\kappa_{5}^{4}}
-\frac{72\kappa_{4}^{2}}{\kappa_{5}^{6}}\sqrt{\frac{\kappa_{4}^{4}}{\kappa_{5}^{4}}
+\frac{1}{3}\kappa_{4}^{2}\rho_{b}+\frac{\kappa_{5}^{4}}{36}W_{0}^{2}}\,\Bigg]^{1/2}.
\end{equation}
To have a simple comparison, we note that in the RSII case this
quantity is given by $\beta^{(RSII)}=
a_{0}\sqrt{\rho_{b}^{2}/W_{0}^{2}-1}$ (see [9] for instance ).
Inserting equation (40) into (16) leads to the following
differential equation
\begin{equation}
\big({a'}\big)^{2}=\frac{{\kappa}_{5}^{4}}{36}\big({\beta}^{2}+{a}^{2}\big){W}^{2}({\phi}).
\end{equation}
We take ${\alpha}_{1}={\alpha}_{2}=\frac{{\kappa_{5}}}{\sqrt{3}}$,
so that the superpotential used in subsection $2.4$ can be
simplified to
\begin{equation}
W=\frac{2c}{{\alpha}_1}\cosh{({\alpha}_1{\phi})}=\frac{c\sqrt{3}}{\kappa_{5}}\Bigg[\Bigg(\frac{{a}_*}{a}\Bigg)^{2}
+4\Bigg]^{\frac{1}{2}}\,.
\end{equation}
Therefore the differential equation has a general solution of the
form ( see [17] for instance )
\begin{equation} a^{2}=A\cosh{\mu y}+B\sinh{\mu y}+C,
\end{equation}
where
\begin{equation}
\mu=\frac{2c\kappa_{5}}{\sqrt{3}}
\end{equation}
and the coefficients $A$, $B$ and $C$ are functions of time.

The  $Z_{2}$ symmetry across the brane imposes the relations
$A_{+}=A_{-}=\bar{A}$ and $B_{+}=B_{-}=\bar{B}$ between these
coefficients on the two sides of the brane [17]. From equations (12)
and (13) we can find the following relations between coefficients
\begin{equation}
\frac{\bar{B}\mu}{\bar{A}+C}=-\frac{\kappa_{5}^{2}}{3}\rho_{b}+
\frac{\kappa_{5}^{2}}{\kappa_{4}^{2}}\frac{\dot{a}_{0}^{2}}{a_{0}^{2}}
\end{equation}
and
\begin{equation}
2\frac{\dot{\bar{B}}\mu}{\dot{\bar{A}}+\dot{C}}-\Big(\frac{\beta
\kappa_{5}^{2}
W_{0}}{3}\Big)^{2}\Big(\frac{\bar{B}\mu}{(\dot{\bar{A}}+\dot{C})^{2}}\Big)=\frac{\kappa_{5}^{2}}{3}(3p_{b}+2\rho_{b})
+\frac{\kappa_{5}^{2}}{\kappa_{4}^{2}}\Big[-\frac{\dot{a}_{0}^{2}}{a_{0}^{2}}-2\frac{\dot{a}_{0}\dot{n}_{0}}{a_{0}}+
2\frac{\ddot{a}_{0}}{a_{0}}\Big]\,.
\end{equation}
Using equations (26), (40) and (44) and taking the limit $y=0$, one
can determine the coefficients as follows
\begin{equation}
C=a_{0}^{2}\Bigg\{1-\frac{2}{\mu^{2}}\Bigg[\bigg(\frac{\dot{a_{0}}}{a_{0}}\bigg)^2
-\frac{\kappa_{5}^{4}}{18}W_{0}^{2}
+\frac{\kappa_{5}^{2}}{\kappa_{4}^{2}}\bigg(\frac{\dot{a_{0}}}{a_{0}}\bigg)^2
+\frac{\kappa_{5}^{2}}{3}\rho_{b}\Bigg]\Bigg\}\,,
\end{equation}
\begin{equation}
\bar{A}=\frac{2a_{0}^{2}}{\mu^{2}}\Bigg[\bigg(\frac{\dot{a_{0}}}{a_{0}}\bigg)^2
-\frac{\kappa_{5}^{4}}{18}W_{0}^{2}+\frac{\kappa_{5}^{2}}{\kappa_{4}^{2}}\bigg(\frac{\dot{a_{0}}}{a_{0}}\bigg)^2
+\frac{\kappa_{5}^{2}}{3}\rho_{b}\Bigg]\,,
\end{equation}
\begin{equation}
\bar{B}=\frac{a_{0}^{2}}{\mu}\Bigg[-\frac{\kappa_{5}^{2}}{3}\rho_{b}+
\frac{\kappa_{5}^{2}}{\kappa_{4}^{2}}\bigg(\frac{\dot{a}_{0}}{a_{0}}\bigg)^{2}\Bigg]\,.
\end{equation}
Substituting these coefficients into the general solution (44), we
obtain the following expression for the bulk behavior of the scale
factor
$${a}^2=a_{0}^{2}\Bigg[\frac{2}{\mu^{2}}\Bigg(\bigg(\frac{\dot{a_{0}}}{a_{0}}\bigg)^2
-\frac{\kappa_{5}^{4}}{18}W_{0}^{2}
+\frac{\kappa_{5}^{2}}{\kappa_{4}^{2}}\bigg(\frac{\dot{a_{0}}}{a_{0}}\bigg)^2
+\frac{\kappa_{5}^{2}}{3}\rho_{b}\Bigg)\bigg(\cosh\big(\mu{y}\big)
-{1}\bigg)\Bigg]$$
\begin{equation}
+{a}^2_{0}\Bigg[1-\frac{1}{\mu}\Bigg(-\frac{\kappa_{5}^{2}}{3}\rho_{b}+
\frac{\kappa_{5}^{2}}{\kappa_{4}^{2}}\bigg(\frac{\dot{a}_{0}}{a_{0}}\bigg)^{2}\Bigg)\sinh{(\mu
|y|)}\Bigg]\,.
\end{equation}

As we mentioned previously, $W\propto a^{-1}$ and therefore, the
singularities in this model occur if $|W|\rightarrow\infty$. In
figure 8, we plotted the scale factor versus $\rho_{b}$ and $y$.
This figure shows that for some values of $\rho_{b}$ and $y$, there
are points in the parameter space that the scale factor vanishes.
The position of these singularities varies with cosmic time on the
brane. At these points, $\phi'$, $\dot{\phi}$ and $W$ all are
singular and is so $T_{Ab}$. Thus the singularities at $a^{2}=0$ are
naked curvature singularities. Nevertheless, if we have a compact
bulk or more other branes at suitable distances from our brane, we
can avoid these singularities. Also, if we have the solutions with
bounded $W$, it is possible to elusion from the curvature
singularities. Note that in ploting figure 8, we have set
$W_{0}=2\rho_{b}$.
\begin{figure}[htp]
\begin{center}\includegraphics{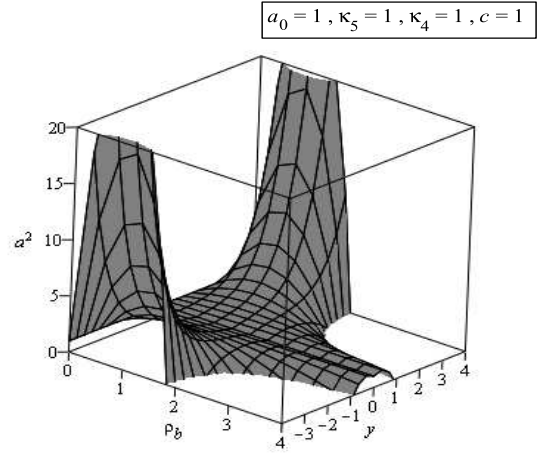} \vspace{6.5cm}
\end{center}
\caption{\small {Evolution of the bulk scale factor. }}
\end{figure}

If we take ${\alpha}_{1}=-{\alpha}_{2}=\frac{\kappa_{5}}{\sqrt{3}}$,
the differential equation (42) has a general solution of the form
\begin{equation}
a^{2}=Ay^{2}+By+C.
\end{equation}
Following the same procedure as above, one can find the following
expression for the bulk scale factor
$${a}^2=a_{0}^{2}\Bigg[1-\frac{1}{\mu}\Bigg(1+\bigg(\frac{\dot{a_{0}}}{a_{0}}\bigg)^2
-\frac{\kappa_{5}^{4}}{18}W_{0}^{2}+\frac{\kappa_{5}^{2}}{\kappa_{4}^{2}}
\bigg(\frac{\dot{a_{0}}}{a_{0}}\bigg)^2+\frac{\kappa_{5}^{2}}{3}\rho_{b}\Bigg)\Bigg(-\frac{\kappa_{5}^{2}}{3}\rho_{b}
+\frac{\kappa_{5}^{2}}{\kappa_{4}^{2}}\bigg(\frac{\dot{a}_{0}}{a_{0}}\bigg)^{2}\Bigg)y\Bigg]$$
\begin{equation}
+{a}^2_{0}\Bigg[\Bigg(\bigg(\frac{\dot{a_{0}}}{a_{0}}\bigg)^2
-\frac{\kappa_{5}^{4}}{18}W_{0}^{2}
+\frac{\kappa_{5}^{2}}{\kappa_{4}^{2}}\bigg(\frac{\dot{a_{0}}}{a_{0}}\bigg)^2
+\frac{\kappa_{5}^{2}}{3}\rho_{b}\Bigg)y^{2}\Bigg]\,.
\end{equation}

Figure 9 shows variation of $a^{2}$ versus $\rho_{b}$ and $y$. As
this figure shows, the bulk scale factor vanishes in some points.
Since $T_{AB}$ is also divergent in these points, there are naked
curvature singularities in the bulk.
\begin{figure}[htp]
\begin{center}\includegraphics{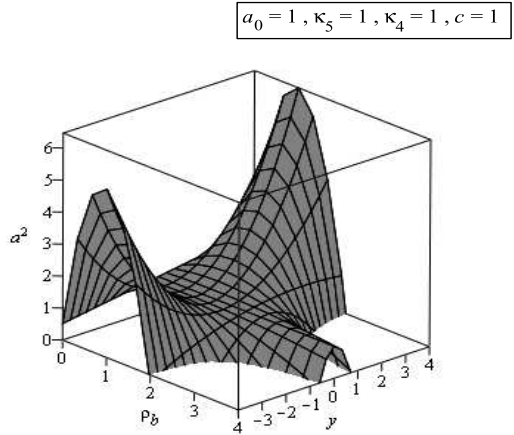} \vspace{6.5cm}
\end{center}
\caption{\small {Evolution of the scalar field. For some values of
$\rho_{b}$ and $y$, it becomes zero. In this points there are naked
singularities.}}
\end{figure}

\section{Bulk phantom scalar field}

\subsection{Field equations}

In this section we consider a DGP-inspired braneworld model that the
bulk contains a phantom scalar field. For a phantom scalar field,
the sign of the kinetic energy term is opposite of the canonical
scalar field case studied in the previous section. The
five-dimensional action for this model can be written as follows
\begin{equation}
S=\int_{bulk}d^{5}x\sqrt{-g}\Big\{\frac{1}{2\kappa_{5}^{2}}R^{(5)}+\frac{1}{2}(\nabla\phi)^{2}-V(\phi)\Big\}
-\int_{brane}d^{4}x\sqrt{-h}\Big(\frac{1}{2\kappa_{4}^{2}}R+\frac{1}{2\kappa_{5}^{2}}[K]+{\cal{L}}_{b}(\phi)\Big)\,.
\end{equation}

Variation of the above action with respect to the phantom scalar
field and also the metric, gives the following equations of motion.
These equations differ from equations (2) and (3) for canonical
scalar field in the sign of the scalar field dependent terms
\begin{equation}
\nabla^{2}\phi=-\frac{dV}{d\phi}-\frac{\sqrt{-h}}{\sqrt{-g}}\frac{d{\cal{L}}_{b}(\phi)}{d\phi}\delta(y)\,,
\end{equation}
\begin{equation}
G_{A}^{B}=\kappa_{5}^{2}\Bigg(-\nabla^{A}\phi\nabla_{B}\phi+\delta_{B}^{A}\Big[\frac{1}{2}(\nabla\phi)^{2}
-V(\phi)\Big]\Bigg)+\delta(y)\kappa_{5}^{2}T_{B}^{A(brane)}\,.
\end{equation}

In the presence of the phantom scalar field, the Gauss-Codacci
junction conditions are the same as for the canonical scalar field
case. Also, if we assume the metric to be as (9), the jump
conditions and the energy-momentum conservation equation, are the
same as given in section 2.1.

In order to find some solutions of the five dimensional field
equations in this case, we use the quantity $F$ and the component of
the Einstein tensor used in section 2.2. By using equation (16), for
special class of solutions with $\phi=\phi(a)$ and so $F=F(a)$,
equations (17) and (18) reduce to
\begin{equation}
\kappa_{5}^{2}V(\phi)+\frac{\kappa_{5}^{2}}{2}
F(a\frac{d\phi}{da})^{2}+\Bigg\{\frac{3\kappa_{5}^{2}}{\kappa_{4}^{2}}\bigg(\frac{\dot{a}}{a}\bigg)^{2}
+\frac{3\kappa_{5}^{2}}{\kappa_{4}^{2}}\bigg(\frac{k}{a^{2}}\bigg)+\kappa_{5}^{2}\,\rho_{b}
\Bigg\}\delta(y)+6F +\frac{3}{2}a\frac{dF}{da}-\frac{3k}{a^{2}}=0\,.
\end{equation}
Also the field equation (55) reduces to
\begin{equation}
\nabla^{2}\phi=a^{2}F\Big(\frac{d^{2}\phi}{da^{2}}+\frac{1}{a}\frac{d\phi}{da}\Big)
+\Big[(G^{0}_{\,0}+G_{\,5}^{5})\frac{a}{3}+\frac{2k}{a}\Big]\frac{d\phi}{da}=-
\frac{d V}{d\phi}-\delta(y)\frac{\sqrt{-h}}{\sqrt{-g}}\frac{\delta
{\cal{L}}_{b}(\phi)}{\delta\phi}\,.
\end{equation}
By using $G_{AB}=\kappa^2 T_{AB}$ and equation (58), equation (57)
gives
$$F(a\frac{d}{da})^{2}\phi+\Bigg\{-\frac{2a}{3}\Big[\frac{\kappa_{5}^{2}}{2}
F(a\frac{d\phi}{da})^{2}+\Big\{\frac{3\kappa_{5}^{2}}{\kappa_{4}^{2}}\big(\frac{\dot{a}}{a}\big)^{2}
+\frac{3\kappa_{5}^{2}}{\kappa_{4}^{2}}\big(\frac{k}{a^{2}}\big)+\kappa_{5}^{2}\,\rho_{b}
\Big\}\delta(y)+6F
+\frac{3}{2}a\frac{dF}{da}\Big]+\frac{5k}{a}\Bigg\}\frac{d\phi}{da}$$
\begin{equation}
+\frac{d V}{d\phi}+\delta(y)\frac{\sqrt{-h}}{\sqrt{-g}}\frac{\delta
{\cal{L}}_{b}(\phi)}{\delta\phi}=0.
\end{equation}

This is the effective field equation for a bulk phantom scalar field
in the DGP setup. In which follows, we present some
supergravity-style solutions for this effective field equation.

\subsection{Supergravity-style solutions}

In this subsection, to present some solutions of the field
equations, we use a supergravity-style potential introduced earlier
in section 3.2. Assuming $k=0$, the field equations (57) and (59)
are satisfied if
\begin{equation}
F=-\frac{\kappa_{5}^{4}}{36}W^{2},
\end{equation}
\begin{equation}
a\frac{d\phi}{da}=-\frac{3}{\kappa_{5}^{2}W}\frac{dW}{d\phi}.
\end{equation}
respectively. The Friedmann equation on the brane now takes the
following form
\begin{equation}
\frac{\dot{a}^{2}_{0}}{a_{0}^{2}}=\frac{1}{3}\kappa_{4}^{2}\rho_{b}+\frac{2\kappa_{4}^{4}}{\kappa_{5}^{4}}
-\frac{2\kappa_{4}^{2}}{\kappa_{5}^{2}}\sqrt{\frac{\kappa_{4}^{4}}{\kappa_{5}^{4}}
+\frac{1}{3}\kappa_{4}^{2}\rho_{b}-\frac{\kappa_{5}^{4}}{36}W_{0}^{2}}\,\,.
\end{equation}
Also, time variation of the phantom scalar field now is given as
follows
\begin{equation}
\dot{\phi}^{2}=\Bigg(\frac{1}{3}\kappa_{4}^{2}\rho_{b}+\frac{2\kappa_{4}^{4}}{\kappa_{5}^{4}}-
\frac{2\kappa_{4}^{2}}{\kappa_{5}^{2}}\sqrt{\frac{\kappa_{4}^{4}}{\kappa_{5}^{4}}+\frac{1}{3}\kappa_{4}^{2}\rho_{b}
-\frac{\kappa_{5}^{4}}{36}W_{0}^{2}}\Bigg)\Bigg(\frac{9}{\kappa_{5}^{4}W^{2}_{0}}\Bigg)
\Bigg(\frac{dW_{0}}{d\phi}\Bigg)^{2}.
\end{equation}
We note that the energy conservation equation is the same as given
by equation (29).

\subsection{A phantom field superpotential}

As discussed in Ref. [19], the curvature of the universe grows
toward infinity within a finite time in the universe dominated by a
phantom fluid. In the case of a phantom scalar field, this Big Rip
singularity may be avoided if the potential has a maximum. Here we
consider the following form of the superpotential which has the
mentioned property
\begin{equation}
W(\phi)={W}_{0}\Bigg(\cosh\big(\frac{{\alpha}{\phi}}{{m}_{pl}}\big)\Bigg)^{-1}
\end{equation}
This superpotential has a maximum at $\phi=0$ and tends to zero when
the phantom scalar field grows to infinity (see figure 8). The
corresponding potential obtained from equation (23) is
\begin{equation}
V(\phi)=\frac{1}{2}{{W}_{0}^{2}}\Bigg[\frac{1}{4}\Big(\frac{{\alpha}}{m_{pl}}\Big)^{2}
\Big(\tanh\big(\frac{{\alpha}{\phi}}{{m}_{pl}}\big)\Big)^{2}-\frac{{\kappa}_{5}^{2}}{3}\Bigg]
\Bigg(\cosh{\big(\frac{{\alpha}{\phi}}{{m}_{pl}}\big)}\Bigg)^{-2}\,.
\end{equation}
The solution of equation (61) when $W$ is given as (64) is as
follows
\begin{equation}
\ln{\Big(\frac{a}{a_{*}}\Big)}=\frac{\kappa_{5}^{2}}{3}\Big(\frac{m_{pl}}
{\alpha}\Big)^{2}\ln\Big(\sinh(\frac{\alpha\phi}{m_{pl}})\Big)\,,
\end{equation}
where $a_{*}$ is an arbitrary constant.
\begin{figure}[htp]
\begin{center}\includegraphics{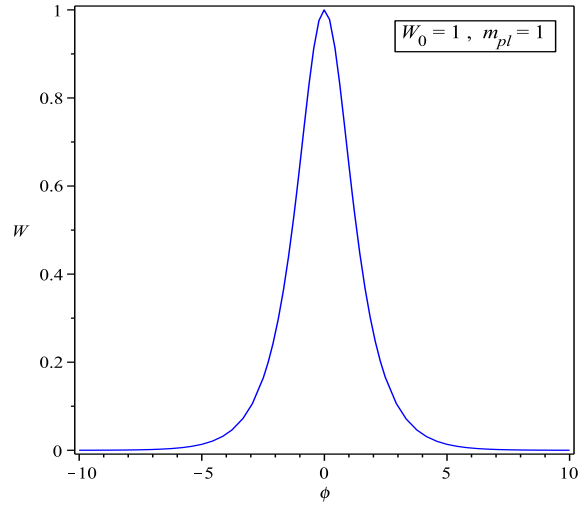} \vspace{6.5cm}
\end{center}
\caption{\small {Evolution of the phantomic superpotential (64) with
respect to the phantom scalar field. This superpotential has a
maximum at $\phi=0$.}}
\end{figure}
This relation shows that for a bulk phantom scalar field, the result
of the superpotential approach for evolution of the scalar field
versus the scale factor, is different for even and odd values of
$\alpha$. For odd values of $\alpha$, scalar field varies from zero
to $+\infty$ as scale factor grows to infinity. But, for even values
of $\alpha$, as scale factor goes to infinity, scalar field varies
from zero to both $\pm\infty$.

\begin{figure}[htp]
\begin{center}\includegraphics{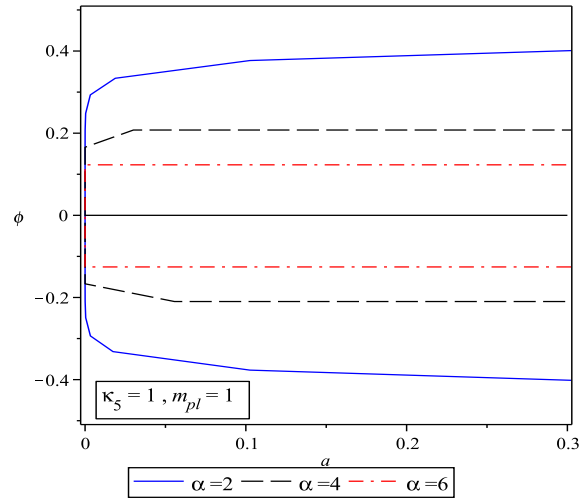} \vspace{6.5cm}
\end{center}
\caption{\small {Evolution of the phantom scalar field versus the
scale factor for even values of $\alpha$.}}
\end{figure}

\begin{figure}[htp]
\begin{center}\includegraphics{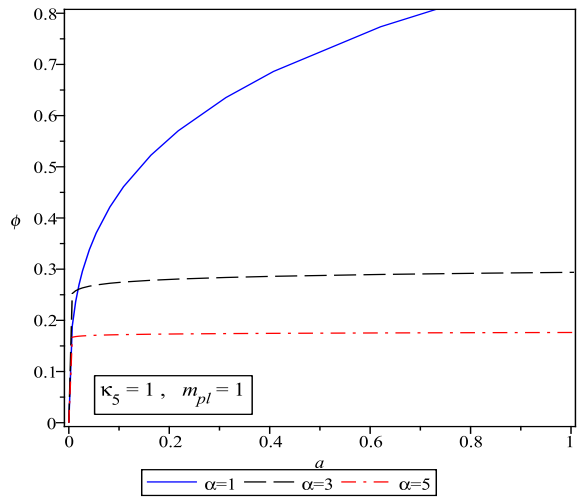} \vspace{6.5cm}
\end{center}
\caption{\small {Evolution of the phantom scalar field with respect
to the scale factor for odd values of $\alpha$.}}
\end{figure}

As figure 8 shows, $W$ has its maximum value at $\phi=0$. In this
point, since $\dot{W_{0}}=0$ too, the right hand side of equation
(29) vanishes and we have the energy conservation on the brane.\\

$\clubsuit$ When $\phi$ goes to $+\infty$,\, $W$ decreases until it
tends to zero when scalar field reaches infinity. During
this course, $\frac{\dot{W}}{W}<0$. This implies that:\\

\textit{\textbf{A}:} For $\rho_{b}>X$, energy leaks off the brane as
scale factor increases and the universe expands. This leakage of
energy continues until the scale factor reaches infinity and $W$
tends to zero. After that there is no leakage of energy-momentum off
the brane.\\

\textit{\textbf{B}:} However, if $\rho_{b}<X$, the situation is
different. Since $\frac{\dot{W}}{W}<0$, so the right hand side of
the conservation equation is positive. This means that energy is
sucked onto the brane whereas the universe expands. This suction
persists until scale factor and the scalar field reach infinity.
After that the suction of energy onto the brane will be
stopped.\\

$\clubsuit$ As $\phi$ goes from $0$ to $-\infty$,\, $W$ decreases
until it tends to zero when scalar field reaches infinity. During
this course, $\frac{\dot{W}}{W}>0$. This implies
that:\\

\textit{\textbf{A}:} For $\rho_{b}<X$, energy leaks off the brane as
scale factor increases and the universe expands. This leakage of
energy continues until $W$ tends to zero. After that there is no
leakage of energy-momentum off the brane.\\

\textit{\textbf{B}:} However, if $\rho_{b}>X$, the right hand side
of the conservation equation is positive. This means that energy is
sucked onto the brane whereas the universe expands. This suction
persists until scale factor and the scalar field reach infinity.
After that the suction of energy onto the brane will be
stopped.\\

We note that since in this case $W$ tends to zero at late time, the
model with bulk scalar field with superpotential as given by
equation (64), is not a viable DGP-like cosmology.

\subsection{Bulk solutions with phantom field}
To find the $y$ dependence of the metric (or ${\phi}$), we proceed
the method used in section 5.2. As before, we assume the fifth
dimension is static ($\dot{b}=0$). When ${\phi}={\phi}(a)$, we have
\begin{equation}
{G}_{05}=3\Bigg(\frac{n'\dot{a}}{na}+\frac{\dot{b}{a'}}{ba}-\frac{\dot{a}'}{a}\Bigg)
={\kappa}_{5}^{2}{T}_{05}={-\kappa}_{5}^{2}{a'}\dot{a}\Bigg(\frac{d{\phi}}{da}\Bigg)^{2}.
\end{equation}
So, we can find the expression of $n$ in terms of $a$
\begin{equation}
\frac{\dot{a}}{n}=\beta(t)\frac{{\kappa}_{5}^{2}}{6}
\exp{\Bigg[\frac{{\kappa}_{5}^2}{3}\int{a\Bigg(\frac{d{\phi}}{da}\Bigg)^2{da}}\Bigg]}\,.
\end{equation}
As before, $\beta$ has no $y$-dependence. If $V$ has a
supergravity-like form as (23), one can simplify the above equation
to
\begin{equation}
\frac{\dot{a}}{n}=-\beta(t)\frac{{\kappa}_{5}^2}{6}W({\phi}) .
\end{equation}
For ${n}_{0}=1$, equation (62) gives
\begin{equation}
\beta=-\frac{{a}_{0}}{{W}_{0}}\Bigg[\frac{12{\kappa}^{2}_{4}{\rho}_{b}}{{\kappa}^{4}_{5}}
+\frac{72{\kappa}^{4}_{4}}{{\kappa}^{8}_{5}}
-\frac{72{\kappa}^{2}_{4}}{{\kappa}^{6}_{5}}\sqrt{\frac{{\kappa}^{4}_{4}}{{\kappa}^{4}_{5}}
+\frac{{\kappa}^{2}_{4}{\rho}_{b}}{3}-\frac{{\kappa}^{4}_{5}}{36}{{W}_{0}^{2}}}\Bigg]^{\frac{1}{2}}\,.
\end{equation}
By substituting $(69)$ into equation $(16)$, we find
\begin{equation}
\big({a'}\big)^{2}=\frac{{\kappa}_{5}^{4}}{36}\big({\beta}^{2}-{a}^{2}\big){W}^{2}({\phi}).
\end{equation}

Following the same procedure as adopted in subsection 2.5, and by
redefinition of $W$ as equation (64), we find the following
expression for the bulk behavior of the scale factor
$${a}^2=a_{0}^{2}\Bigg[\frac{2}{\mu^{2}}\Bigg(\bigg(\frac{\dot{a_{0}}}{a_{0}}\bigg)^2
-\frac{\kappa_{5}^{4}}{18}W_{0}^{2}
+\frac{\kappa_{5}^{2}}{\kappa_{4}^{2}}\bigg(\frac{\dot{a_{0}}}{a_{0}}\bigg)^2
+\frac{\kappa_{5}^{2}}{3}\rho_{b}\Bigg)\bigg(\cosh\big(\mu{y}\big)-{1}\bigg)\Bigg]$$
\begin{equation}
+{a}^2_{0}\Bigg[1-\frac{1}{\mu}\Bigg(-\frac{\kappa_{5}^{2}}{3}\rho_{b}+
\frac{\kappa_{5}^{2}}{\kappa_{4}^{2}}\bigg(\frac{\dot{a}_{0}}{a_{0}}\bigg)^{2}\Bigg)\sinh{(\mu
|y|)}\Bigg]\,.
\end{equation}
We note that for a bulk phantom scalar field with a bounded
superpotential, as given by (64), the model has no singularity.

\section{Summary and Discussion}

In this paper, we have studied cosmological dynamics in a DGP setup
with a bulk canonical/phantom scalar field. Bulk scalar field is
motivated in several context: to stabilize the distance between two
branes in the Randall-Sundrum two-brane model, a way to solve the
cosmological constant problem, and inflation driven by bulk scalar
field without inflaton on the brane. Generally, the scalar field
living in the bulk affects the cosmological dynamics on the brane
considerably. The evolution of this field has interesting
cosmological effects and can give rise self-acceleration and
phantom-like phase even in the normal DGP branch of the model in
some appropriate situations. In this paper, we have generalized the
work of Ref. [9] to the DGP setup. We considered an extension of the
DGP scenario that the bulk is non-empty and contains a canonical or
phantom scalar field. Since the self-accelerating DGP branch has
ghost instabilities, we restricted our study to the normal DGP
branch of the model. The bulk equations of motion are derived and
some special classes of the solutions are presented. We determined
also the evolution of the brane when the potential of the scalar
field takes a supergravity-like form. Some clarifying examples along
with numerical analysis of the model parameters space are presented
in each step. The importance of this work is that bulk scalar field
in the DGP setup was not studied in supergravity-style analysis.
Also our detailed study in this framework has revealed some yet
unexplored aspects of cosmological dynamics of the bulk scalar field
in DGP setup. We have extended this study to the case that the bulk
contains a phantom scalar field too. Our strategy to perform the
mentioned analysis was as follows:

First of all, from a five-dimensional action for a DGP-inspired
braneworld model with a bulk canonical scalar field, we found the
bulk equations of motion, jump conditions and the Gauss-Codacci
equations. Then, using these equations, we achieved the effective
energy-momentum conservation equation on the brane. We saw that
because of the presence of the time-dependent bulk scalar field and
$\phi$-dependent couplings in the standard model Lagrangian, the
right hand side of the continuity equation is non-zero, showing the
amount of energy non-conservation (due to bulk-brane energy-momentum
transfer) of the matter fields on the brane. To obtain a special
class of solutions for a DGP braneworld cosmology with a bulk scalar
field, we used the methods presented in Refs. [9,16,17]. Using that
method and introducing the quantity $F$ as a function of $t$ and
$y$, we reduced the original partial differential field equations to
an ordinary differential equation. In order to generate some
solutions of the field equations, we introduced a special
supergravity-style potential $V(\phi)$, including the superpotential
$W$. With this supergravity-style potential, we derived the energy
conservation equation on the brane in terms of $W$. In our model,
due to the presence of two terms on the right hand side of the
conservation equation (29), there was new possibilities with
different cosmological implications. If the right hand side of the
conservation equation becomes negative, the energy leaks off the
brane. However, if the right hand side of this equation becomes
positive, there is energy suction onto the brane. For vanishing
right hand side of the conservation equation, energy is conserved on
the brane. We considered some specific examples of superpotential
and discussed their cosmological implications. Evolution of the
scalar filed versus the scale factor, and evolution of $W$ in terms
of the scalar field are discussed fully to study the status of the
conservation equation on the brane. Assuming the simplest
generalization of the brane energy density $(\rho_{b}=W_{0}\rho)$,\,
implies that if $W_{0}$ tends to $+\infty$ at late time, the
cosmological evolution on the brane is not generally compatible with
observations. On the other hand, for $W_{0}=0$, the result is not
compatible with observation too. Only for a constant $W_{0}$ we find
a viable DGP-like cosmology in this simple generalization. So, we
found in those examples that the case with $\alpha_{2}>0$ and $s=+1$
(subsection 2.4.1) is a viable DGP-like cosmology. Of course in the
mentioned case, the situation that the energy leaks of the brane
during universe expansion is more favorable. We also determined the
bulk behavior of the metric. We saw that since $W$ is not bounded,
the naked curvature singularities in the bulk are present. However,
with a compact bulk or existence of more other branes at suitable
distances from our brane, we can avoid these singularities. Also,
for the solutions with bounded $W$, it is possible to elusion from
the curvature singularities. We continued our treatment by
considering the bulk phantom scalar field. From the five-dimensional
action for a DGP-inspired braneworld model with a bulk phantom
scalar field, we found the bulk equations of motion. In the presence
of the bulk phantom scalar field, the Gauss-Codacci junction
conditions and also the energy-momentum conservation equation, are
the same as for the canonical scalar field case. Introducing
quantity $F$ in terms of metric components as before, we achieved
the effective field equation for a bulk phantom scalar field in the
DGP setup. To present some solutions of the field equations, we used
a supergravity-style potential introduced earlier in the case of
canonical scalar field but with a new superpotential. We considered
the evolution of the scalar field versus the scale factor and the
evolution of $W$ versus $\phi$ in order to study the status of the
conservation equation on the brane. Since in this case $W$ tends to
zero at late time, this case is not a viable DGP-like cosmology.
Also, we determined the bulk behavior of the metric. Since in this
case $W$ is bounded, there is no singularity in the brane.\\

Finally the following issues are important to note:

\begin{itemize}

{\item To be a cosmologically viable scenario, this model with bulk
scalar field should explain at least the late-time cosmic speed-up
on the brane. As we have shown, this model accounts for cosmic
speed-up on the brane in some specific situations. For instance, the
case with $\alpha_{2}>0$ and $s=+1$ in subsection 2.4.1 gives a
viable DGP-like cosmology explaining late-time cosmic acceleration.
Also it is possible to realize an effective phantom-like
prescription on the brane without need to phantom matter in the
same way as has been done in Ref. [18] for a canonical scalar
field on the brane.}\\

{\item  As has been mentioned by Flanagan \emph{et al.} (in Ref.
[8]), the principal problem associated with the introduction of a
bulk scalar field in models such as the present one is that generic
stationary solutions contain timelike curvature singularities in the
bulk at finite distances from the branes. One approach to overcome
this problem is to simply orbifold or otherwise compactify the fifth
dimension in so that the singularity is never encountered. A second
approach is to carefully choose the scalar field potential in such a
way that the occurrence of singularities is prevented. Here we have
adopted the second strategy by choosing bounded superpotentials. For
instance, with a bulk phantom scalar field with a bounded
superpotential, the model has no singularity.}\\

{\item  As has been shown in Refs. [12] and [13], in the presence of
a bulk scalar field, realization of the inflation is possible
without inflaton field on the brane. In fact, inflation can be
driven just by a bulk scalar field. In our setup, it is possible to
realize inflation in the same way as has been adopted in Refs. [12]
and [13]. In fact, the late-time behavior of the bulk scalar field
can be treated by analyzing the property of a retarded Green
function. Including the lowest order back-reaction to the geometry,
this late-time behavior can be well approximated by an effective
four dimensional scalar field on the brane. As has been shown by
Himemoto \emph{et al.} in Refs. [12,13], the mapping to the
four-dimensional effective theory is given by a simple scaling of
the potential with a re-definition of the field. This effective
four-dimensional field can drive inflation on the brane. This issue
is under study and will be addressed in one of our forthcoming report.}\\
\end{itemize}

{\bf Acknowledgement}\\

We are very grateful to an anonymous referee for his/her insightful
comments.

\end{document}